\documentclass[12pt]{article}
\usepackage{epsfig}
\usepackage{latexsym}
\input{amssymb.sty}

\def\frac#1#2{{{#1}\over{#2}}}

\newcommand{\bq}{\begin{equation}}
\newcommand{\dq}{\end{equation}}

\begin{document}
\baselineskip 24pt
\begin{center}
{\large\bf Mass and mixing of fermions from a rotating mass 
matrix}\footnote{Updated version of a talk given by the first author 
    at the International
    Symposium on Frontiers of Science - In Celebration of the 80th
    Birthday of C N Yang, June 2002, Beijing.}\\
\ \\
TSOU Sheung Tsun\\
and\\
Jonathan S.\ PALMER\\
Mathematical Institute, Oxford University\\
24--29 St.\ Giles', Oxford OX1 3LB\\
United Kingdom\\
tsou\,@\,maths.ox.ac.uk \quad palmerj\,@\,maths.ox.ac.uk
\end{center}

{\bf Abstract}\\
\baselineskip 14 pt

We summarize the results showing that all existing data on mixing 
between up and down
fermion states (i.e.\ CKM matrix and neutrino oscillations) and on the
hierarchical quark and lepton mass ratios between generations are
consistent with the two phenomena being both consequences of a mass
matrix rotating in generation space with changing energy scale.  
Thus the rotation of the mass matrix can be traced over some 14
orders of magnitude in energy from the mass scale of the $t$-quark at
175 GeV to below that of the atmospheric neutrino at 0.05~eV.

\clearpage

By a rotating mass matrix we mean that the mass matrix of a fermion
species, for example any one of $U$ the up quarks ($t,c,u$), $D$ the
down quarks ($b,s,c$), $L$ the charged leptons ($\tau,\mu,e$) and 
$N$ the neutrinos ($\nu_\tau,\nu_\mu,\nu_e$), has not only
scale-dependent eigenvalues but also scale-dependent eigenvectors.
This last property means that the eigenvectors change
orientation (or rotate) in generation space. 

This is not just a hypothetical situation, because already in the
Standard Model the renormalization group equation for the $U$ mass
matrix has a term \cite{rge}
\bq
16 \pi^2 \frac{dU}{dt} = -\frac{3}{2}D D^{\dagger} U + \ldots,
\dq
where the factor $D^{\dagger} U = V$ is the CKM quark mixing matrix.  Now
this matrix is experimentally found to be non-diagonal, meaning that
even if $U$ is diagonal at a certain scale, it will become nondiagonal
as a result of running to a different scale, at which when it is
re-diagonalized the eigenvector will have rotated from its previous
orientation.   However, given that the off-diagonal elements of the
CKM matrix are small, this effect can usually be neglected in practice
in the Standard Model.   This may very well not be the case in
extensions of the Standard Model, where other forces may contribute to
the mass matrix rotation to an observable extent.

In this note we shall turn the argument around and seek experimental
evidence of a rotating mass matrix, in the manner we shall now detail.

There are two well established experimental facts which the Standard
Model does not explain, namely fermion mass hierarchy and fermion mass
mixing.  The masses of the up and down quarks, and the charged
fermions differ by more than 1--2 orders of magnitude as one goes from
one generation to the next \cite{databook}, as can be seen in 
Table~\ref{fermmass}.
\begin{table}[ht]
\begin{tabular}{llccr}
$\hspace*{3.2cm}$& $t$ &$\hspace*{1cm}$&$\hspace*{1cm}$&   $174300\pm 5100$\\
$\hspace*{3cm}$& $c$ & $\hspace*{1cm}$& $\hspace*{1cm}$& $1150-1350$\\
$\hspace*{1cm}$& $u$ & $\hspace*{1cm}$& $\hspace*{1cm}$& $1-5$\\
&&&&\\
$\hspace*{1cm}$& $b$ & $\hspace*{1cm}$&$\hspace*{1cm}$&  $4000-4400$\\
$\hspace*{1cm}$& $s$ & $\hspace*{1cm}$& $\hspace*{1cm}$& $75-300$\\
$\hspace*{1cm}$& $d$ & $\hspace*{1cm}$& $\hspace*{1cm}$& $3-9$\\
&&&&\\
$\hspace*{1cm}$& $\tau$ & $\hspace*{1cm}$& 
$\hspace*{1cm}$& $1777.03\pm 0.3$\\
$\hspace*{1cm}$& $\mu$ & $\hspace*{1cm}$& $\hspace*{1cm}$& $105.658$\\
$\hspace*{1cm}$& $e$ & $\hspace*{1cm}$& $\hspace*{1cm}$& $0.511$\\
&&&&
\end{tabular}
\caption{Fermion masses in MeV.}
\label{fermmass}
\end{table}
The absolute values of the various CKM matrix elements have been
measured to a good accuracy \cite{databook}:
$$
 \left( \begin{array}{lll}
V_{ud} & V_{us} & V_{ub}\\
V_{cd} & V_{cs} & V_{cb}\\
V_{td} & V_{ts} & V_{tb} \end{array} \right) =
\left( \begin{array}{lll} 
       0.9742 - 0.9757 & 0.219 - 0.226 & 0.002 - 0.005 \\
       0.219 - 0.225 & 0.9734 - 0.9749 & 0.037 - 0.043 \\
       0.004 - 0.014 & 0.035 - 0.043 & 0.9990 - 0.9993 \end{array} \right)
$$
The absolute values of the leptonic MNS mixing matrix, on the other
hand, are not so well determined at present, but from neutrino
oscillation experiments \cite{nuexp} we know that the mixing is in general
considerably larger than the corresponding CKM mixing.  The
experimental estimates are:
$$
\left( \begin{array}{lll}
U_{e1} & U_{e2} & U_{e3}\\
U_{\mu 1} & U_{\mu 2} & U_{\mu 3}\\
U_{\tau 1} & U_{\tau 2} & U_{\tau 3} \end{array} \right) =
\left( \begin{array}{ccc}
       \ast & 0.4 - 0.7 & 0.0 - 0.15 \\
       \ast & \ast & 0.56 - 0.83 \\
       \ast & \ast & \ast \end{array} \right)
$$

Since the Standard Model gives no explanation for these two very
remarkable experimental facts, one is tempted to seek other origins
for them, and if one could find a single mechanism which could give
rise to both, it would certainly provide a very interesting and
promising avenue of study \cite{cevidsm}.

We now make the following rotating hypothesis:
\begin{quote}
 The rotating mass matrix gives rise to all observed mixing and lower
generation masses.
\end{quote}

We note as a side remark that this hypothesis is made and implemented 
in the Dualized Standard Model \cite{dsmrefs}, from which we obtain 
good agreement with
experiment\footnote{Specifically, with 3 real parameters fitted to data, 
it gives correctly to within present experimental bounds the following 
measured quantities: the mass ratios $m_c/m_t$, $m_s/m_b$, 
$m_\mu/m_\tau$, all 
9 elements $|V_{rs}|$ of the CKM matrix, plus  the 2 elements
$|U_{\mu3}|$ 
and $|U_{e3}|$ of the MNS matrix.  
It gives further by interpolation sensible though inaccurate estimates for 
the following: the mass ratios $m_u/m_t$, $m_d/m_b$, $m_e/m_\tau$ 
and the solar neutrino angle $U_{e2}$.  Moreover, numerous detailed 
predictions 
have been made, and tested against data, 
in flavour-violation effects over a wide area comprsising 
meson mass differences, rare hadron decays, $e^+ e^-$ collisions, and 
muon-electron conversion in nuclei, with further predictions  
on effects as far apart in energy as neutrinoless double-beta decays in 
nuclei and cosmic ray air showers beyond the GZK cut-off of $10^{20}$ eV
at the extreme end of the present observable energy range.}.

So we start with a mass matrix which is of rank 1, 
and which remains factorized
on running, so that at any scale we have only one non-zero eigenvalue
$m_T$:
\bq
m=m_T \left( \begin{array}{c}
\xi\\\eta\\\zeta \end{array} \right) 
(\xi\ \eta\ \zeta) = m_T \left( \begin{array}{ccc}
\xi^2 & \xi \eta & \xi\zeta\\
\xi \eta & \eta^2 & \eta \zeta \\
\xi \zeta & \eta \zeta & \zeta^2
\end{array} \right),
\label{massmat}
\dq
Thus the whole mass matrix can be encapsulated by one single 
normalized rotating vector $v=(\xi,\eta,\zeta)$
(without loss of generality we may assume $\xi \geq \eta \geq \zeta$).  
Note that the eigenvalue 
$m_T$ depends on the fermion species 
$T = U, D, L, N$.

Care has to be taken in
defining physical masses and states when the mass matrix runs with
scale.  In our case when the mass matrix is given by a single vector
we can adopt the following procedure \cite{physcons} which 
guarantees that:
\begin{itemize}
\item lepton state vectors are always orthogonal,
\item mixing matrix is always unitary,
\item mass hierarchy is automatic.
\end{itemize}

\begin{figure}
\centerline{\psfig{figure=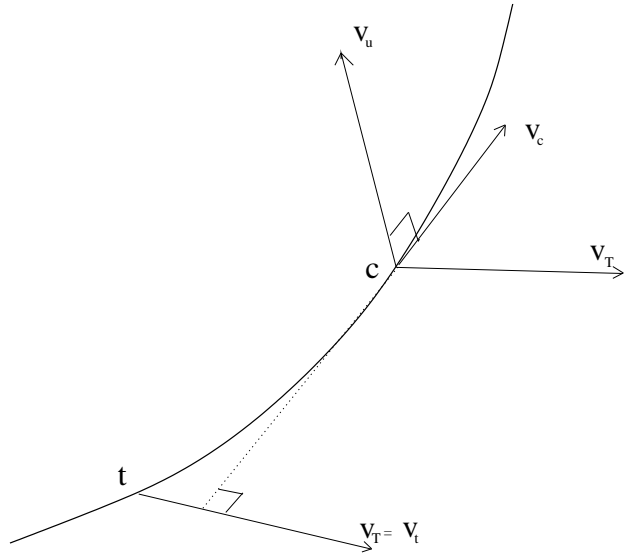,width=0.6\textwidth}}
\caption{The state vectors of the 3 physical states of the $U$ type quark.}
\label{quarkstates}
\end{figure}

First we run $m$ to a scale $\mu$ such that $\mu=m_T(\mu)$; this value
we can reasonably call the mass of the highest generation.  To fix
ideas, let us concentrate on the $U$ type quarks $t,c,u$.  The
corresponding eigenvector $v_t$ is then the state vector of the
$t$ quark.  Having fixed this, we now know that the $c$ and $u$ lie in
the 2-dimensional subspace $V$ orthogonal to $v_t$.  As we go down in
scale the eigenvector $v$ rotates so that $V$ is no longer the null
eigenspace, and the projection of $m$ onto $V$ is a $2 \times 2$
matrix, again of rank 1.  We now repeat the procedure for this
submatrix and determine both the mass and the state of the $c$ quark
(Figure \ref{quarkstates}).
Once the $c$ state is determined, we know the $u$ state as well, as
being the third vector of the orthonormal triad $(t,c,u)$.  The $u$
mass is similarly determined.

\begin{figure}
\centerline{\psfig{figure=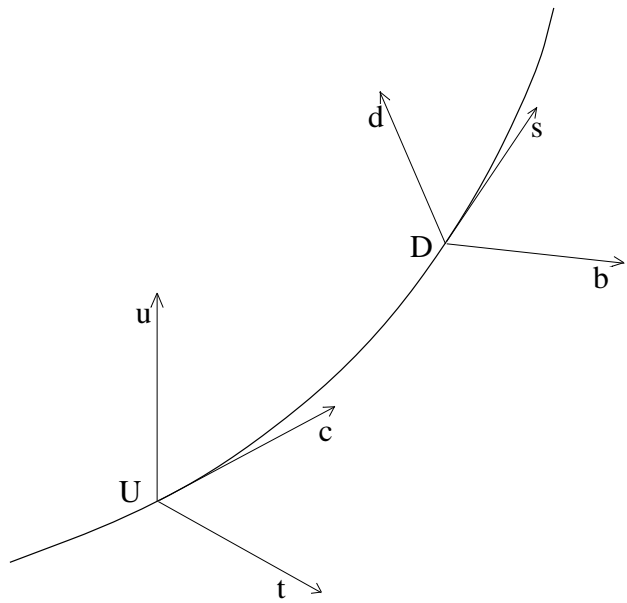,width=0.6\textwidth}}
\caption{Two triads of state vectors for the quarks on the trajectory.}
\label{quarktriads}
\end{figure}

The above procedure can be repeated for the other three types of
fermions\footnote{Neutrinos will need further special treatment which
we shall omit here.   Instead see reference \cite{cevidsm}.}.  
The direction cosines of the two triads $(t,c,u)$ and
$(b,s,d)$, in other words, the 9 inner products, will give us the CKM
matrix, Figure \ref{quarktriads}.
Similarly for the MNS mixing matrix of the leptons.

The above procedure takes a much simpler form when we have only two
generations, and we shall in fact study this case first, as a planar
approximation for the full three generation case, to better illustrate 
the methodology and the results \cite{cevidsm}.

\begin{figure} [ht]
\centering
\input{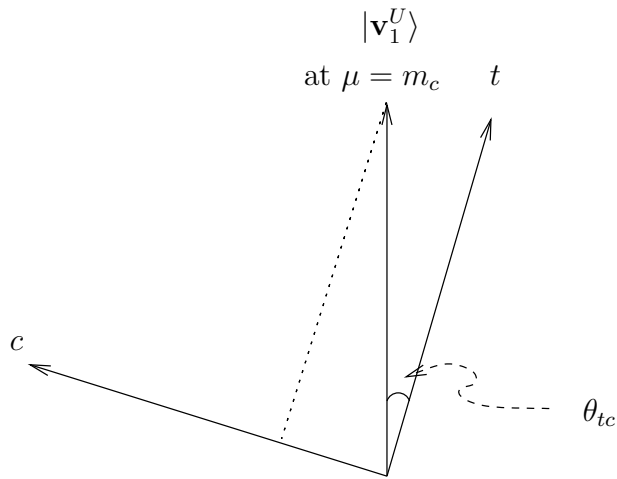}
\caption{Masses for lower generation fermions 
via the ``leakage'' mechanism.}  
\label{planeleak}
\end{figure}

Consider the up quarks first.  When applied 
to this simple case, the procedure detailed above for defining masses 
and state vectors of flavour states gives the state vector $v_t$ 
of $t$ as the single massive eigenstate $v$ of the $U$-quark mass 
matrix at the scale $\mu = m_t$, and the vector $v_c$ as a vector 
orthogonal to $v_t$, as depicted in Figure \ref{planeleak}.
But this should not be interpreted to mean that $c$ has zero mass,
because $m_c$ is given by the nonzero eigenvalue of the mass
submatrix, now only $1 \times 1$, and is hence just the expectation value
of $m$ in the state vector $v_c$, namely 
\bq
m_c = m_t\,\sin^2 \theta_{tc} \neq 0,
\label{planec}
\dq
where $\theta_{tc}$ is the rotation angle between the scales $\mu = m_t$
and $\mu = m_c$.
Similarly, although the mass matrices of the $U$ and $D$ quarks 
are aligned in orientation at all scales, one sees from 
\begin{figure} [ht]
\centering
\input{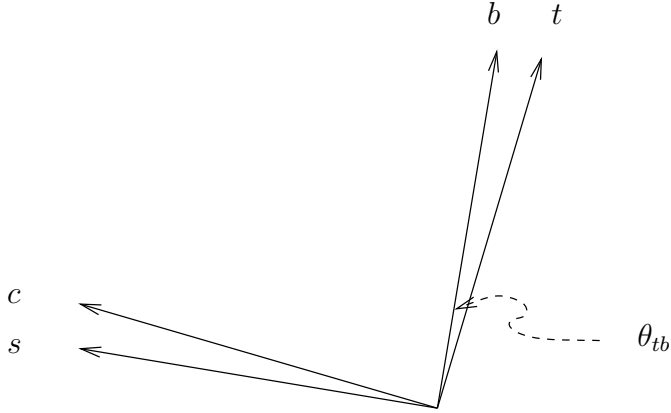}
\caption{Mixing between up and down fermions.}  
\label{planemix}
\end{figure}
Figure \ref{planemix} that by virtue of the rotation of the vector $v$ 
from the scale $\mu = m_t$ to the scale $\mu = m_b$ where the state vectors 
$v_t$ and $v_b$ are respectively defined, the two state vectors 
will not point in the same direction, having rotated by an angle
$\theta_{tb}$ between the two scales:
\bq
V_{tb} = {\bf v}_t.{\bf v}_b = \cos \theta_{tb} \neq 1.
\label{planetb}
\dq
Other fermion species are similar.

\begin{figure}
\centering
\hspace*{-3.5cm}
\input{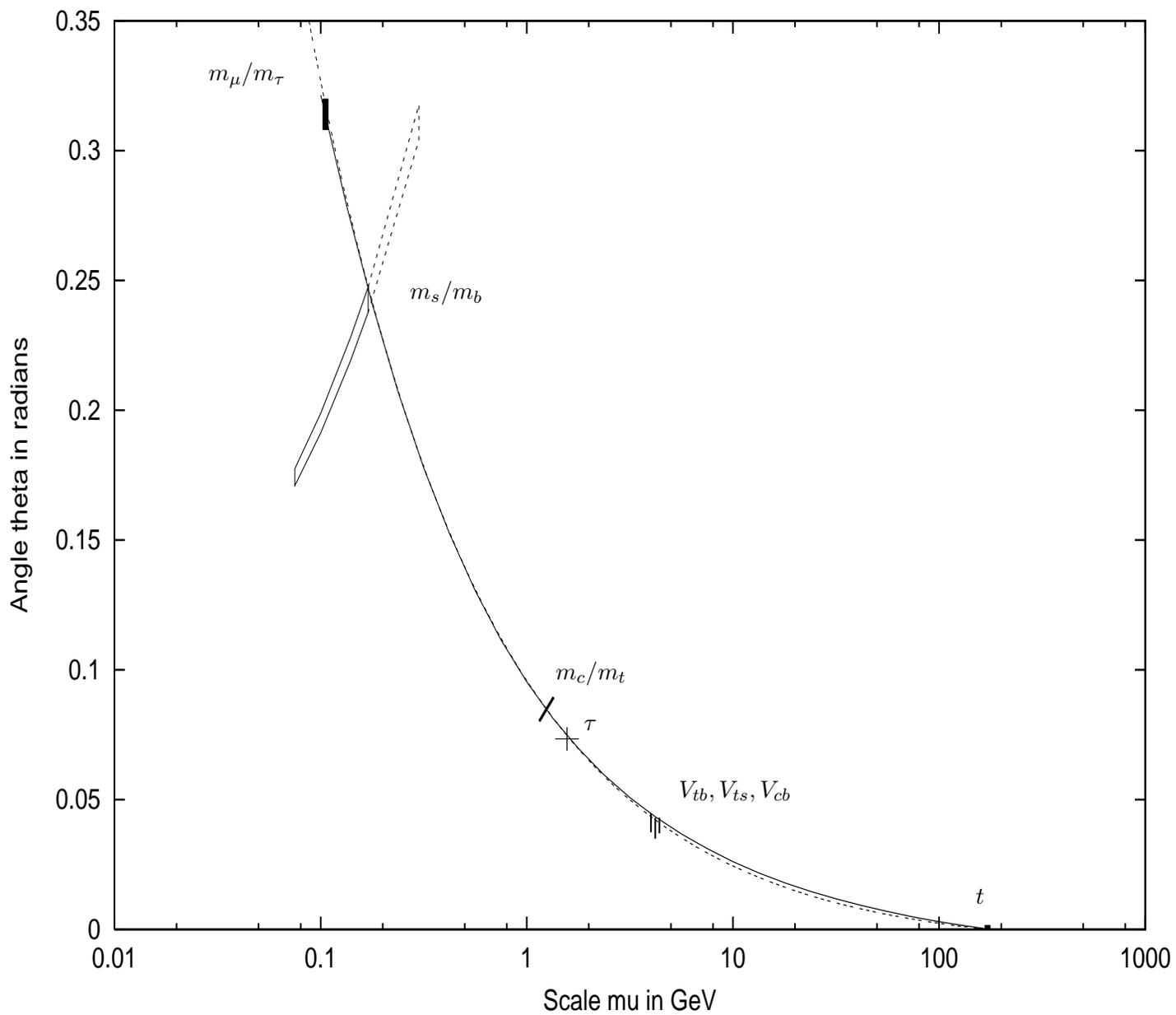}
\caption{The rotation angle changing with scale as extracted from data,
in the planar approximation.}
\label{planero}
\end{figure}

Using (\ref{planec}) and (\ref{planetb}), and Figures \ref{planeleak}
and \ref{planemix}, and similar ones for the other fermions, 
we obtain the following relations:
\bq
V_{tb} =\cos \theta_{tb}, \ |V_{ts}|=|V_{cb}|=\sin \theta_{tb},
\dq
from the mixing, and from the masses
\bq
m_c/m_t = \sin^2 \theta_{tc},\ m_s/m_b = \sin^2 \theta_{bs},\ 
m_\mu/m_\tau = \sin^2 \theta_{\tau\mu}.
\dq
Inputting from data \cite{databook} on mixing
\bq
|V_{tb}| = 0.9990 - 0.9993 \  |V_{ts}| = 0.035 - 0.043,\ |V_{cb}| =
0.037 - 0.043, 
\dq
and on masses for up quarks
\bq
m_t = 174.3 \pm 5.1\ {\rm GeV}, \ \ m_c = 1.15 - 1.35\ {\rm GeV},
\dq
for down quarks
\bq
m_b = 4.0 - 4.4\ {\rm GeV}, \ \ m_s = 75 - 170\ {\rm MeV},
\dq
and for charged leptons
\bq
m_\tau = 1.777\ {\rm GeV}, \ \ m_\mu = 105.66\ {\rm MeV};
\dq
we obtain compatible values for
\bq
\theta_{tb} = 0.0374 - 0.0447,\  0.0350 - 0.0430,\  0.0370- 0.0430
\dq
and 
\bq
\theta_{tc} = 0.0801 - 0.0894,\ \theta_{bs} = 0.1309 - 0.2076,\ 
\theta_{\tau \mu} = 0.2463.
\dq
These values suggest a smooth curve, which will be the case if
these angles are swept out by a rotating vector $v$.   This is clearly
borne out by Figure \ref{planero},
where the dotted curve represents the best-fit to data using MINUIT:
\bq
\theta = \exp (-2.267 -0.509 \ln \mu) -0.0075
\dq
$\mu$ in GeV, with an excellent $\chi^2$ of 0.21 per degree of
freedom.  The solid
curve shown is an earlier calculation \cite{massckm, phenodsm} 
to 1-loop in the Dualized
Standard Model (see Footnote 1).  The two curves are almost
indistinguishable, and hence fit the data points equally well.

The positive result from the two-generation case is encouraging, but
below the scale of roughly the $s$ quark mass, nonplanar effects begin to
be appreciable, as can be estimated by the square of the Cabibbo angle
which gives about 4 \%.  Therefore to go further down the scale it is
necessary  
to study the full three-generation case \cite{cevidsm}.  Writing
\bq
v(\mu) =  (\xi(\mu),\eta(\mu),\zeta(\mu)),
\dq
we shall present the extracted angles by plotting $\eta(\mu),
\zeta(\mu)$ against scale $\mu$.

First we fix the $U$ triad to be
\bq
U \colon (1,0,0),\ (0,1,0), \ (0,0,1).
\dq
Then the $D$ triad is just given by the elements of the CKM matrix:
\bq
D \colon (V_{tb}, V_{cb}, V_{ub}),\ (V_{ts},V_{cs},V_{us}),\ (V_{td},
V_{cd}, V_{ud}).
\dq
We thus obtain the points $t$ and $b$ on the plot (Figure
\ref{3Dplot}).

\begin{figure*}
\vspace*{-3cm}
\centering
\includegraphics[angle=90,scale=1.0]{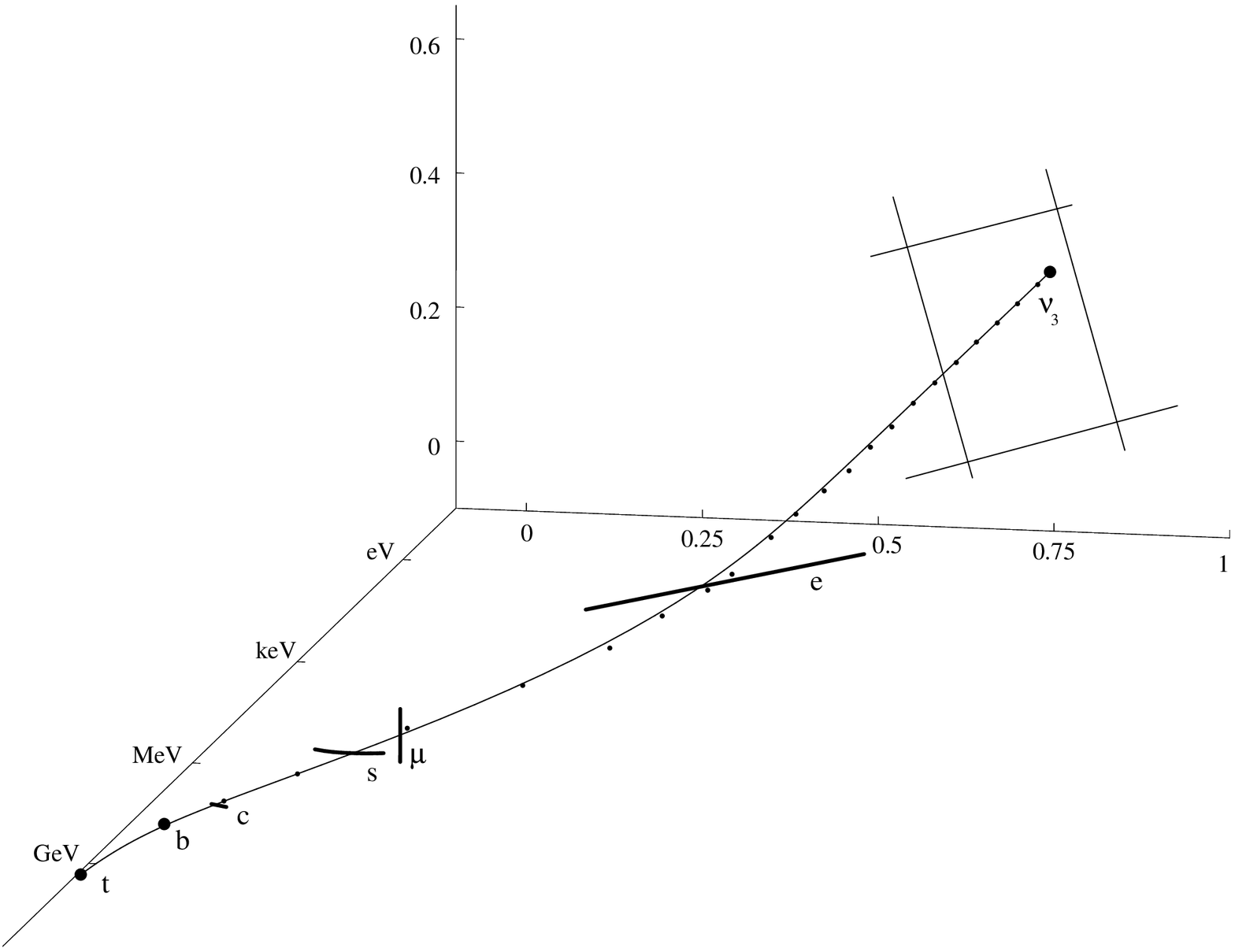}
\end{figure*}

\begin{figure}
\vspace{3cm}
\caption{A plot of the rotating vector $v(\mu)$ with its 
second and third
components, i.e.\ $\eta(\mu)$ and $\zeta(\mu)$, as
functions of $\ln \mu$, $\mu$ being the energy scale, showing a smooth
curve (dotted) through the experimentally allowed regions.   
The solid
curve represents the result of a DSM one-loop calculation from an earlier 
paper \cite{phenodsm} which is seen also to pass through the allowed regions 
except that for the electron $e$.  Below the $\mu$ lepton mass the two
curves are not distinguishable.} 
\label{3Dplot}
\end{figure}

\begin{figure}
\centering
\includegraphics[scale=1.4]{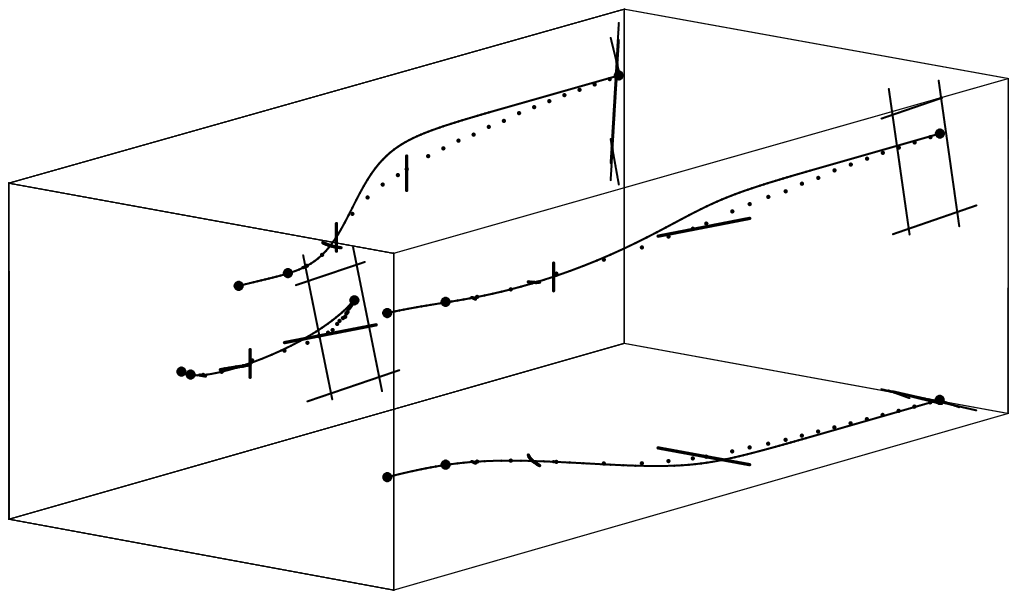}
\caption{The same figure as in Figure~\ref{3Dplot}, seen with its three
projections onto the coordinate planes.}
\label{proj}
\end{figure}

Next we consider $v(\mu=m_c)$.  Since the mass $m_c$ comes from
`leakage' from $m_t$, we can write
\begin{eqnarray}
v (\mu=m_c)&=& \cos \theta_{tc}  v_t + \sin \theta_{tc}
v_c \nonumber \\
{}&=& \sqrt{1-m_c/m_t}  v_t + \sqrt{m_c/m_t}  v_c,
\end{eqnarray}
which gives the point $c$ in Figure \ref{3Dplot}.  Note that the
existence of another disjoint branch of the square root 
does not affect the question of interest
to us here, namely, whether the allowed region is consistent with data
lying on a smooth rotation curve, so long as the first branch already
does, and can therefore be ignored.

Similar considerations give us a line as the allowed region for the
$s$ quark.  We shall leave open the question of the light quarks $u$
and $d$, since they are tightly bound and both the definition and 
the value of their masses
are quite uncertain.

In principle the leptons are not linked to the quarks, but the planar
approximation above (Figure \ref{planero}) and the DSM 
calculations \cite{massckm, phenodsm}
both suggest that they can be put on the same trajectory as the
quarks.   In fact, by putting the $\tau$ point by interpolation
between the $b$ and $c$ points, we can easily continue the rotation 
curve with the leptons, as is again evident from Figure \ref{3Dplot}.
In Figure~\ref{proj} we show the same curve with its three projections 
onto the three planes: $\mu-\eta$, $\mu-\zeta$, and $\eta-\zeta$.

We have not given in any detail on how the actual angles are extracted
from the quark and lepton data, only the theory of it.  Full details
can be found in \cite{cevidsm}.

In conclusion we can say that, using the rotation hypothesis, and all
the fermion mass and mixing data (apart from $m_u$ and $m_d$), we
exhibit a smooth curve in 3-d space, passing through all the allowed
regions, and spanning some 14 orders of magnitude in energy.
Moreover, this curve coincides with the DSM curve \cite{phenodsm}
where we expect it
to do, that is, where the 1-loop approximation is good. 

Most of the work reported here was done in collaboration with Jos\'e 
Bordes and Chan Hong-Mo.

\end{document}